\documentclass{amsart}

\usepackage{amssymb, amsmath, amsthm}
\usepackage{wasysym}
\usepackage{multicol}
\usepackage{newalg}
\usepackage{tikz}
\usepackage{hyperref}

\newtheorem{theorem}{Theorem}

\theoremstyle{remark}

\newtheorem{example}[theorem]{Example}

\newcommand{\alp}{{\rm alph}}
\newcommand{\emp}{\varepsilon}

\newcommand{\E}{{\mathcal E}}

\newcommand{\tE}{{\tt E}}
\newcommand{\tL}{{\tt L}}
\newcommand{\tl}{{\tt l}}
\newcommand{\tr}{{\tt r}}
\newcommand{\tk}{{\tt k}}
\newcommand{\tR}{{\tt R}}
\newcommand{\C}{{\mathcal C}}

\newcommand{\R}{{\mathcal R}}
\newcommand{\bigo}{{\mathcal O}}

\newcommand{\tz}{\ |\ }

\newcommand{\prav}[1]{{\mathbf r}_#1}
\newcommand{\lev}[1]{{\mathbf l}_#1}
\newcommand{\okol}[1]{{\mathbf n}_{#1}}

\newcommand{\comment}[1]{}

\newcommand{\MF}{{\sc MorphicFactorization}}

\newcommand{\tN}{{\tt New}}
\newcommand{\Pos}{{\tt Pos}}

\newcommand{\mytikz}[1]{}

\begin{document}

\title{Complexity of testing morphic primitivity}

\author{Vojt\v ech Matocha}

\author{\v St\v ep\' an Holub}

\address{Department of Algebra, Charles University, Sokolovsk\'a 83, 175 86 Praha, Czech Republic}
\email{vojtech.matocha@bcvsolutions.eu, holub@karlin.mff.cuni.cz}



\begin{abstract}
We analyze the algorithm in \cite{DM}, which decides whether a given word is a fixed point of a nontrivial morphism. We show that it can be implemented to have complexity in $\bigo(m\cdot n)$, where $n$ is the length of the word and $m$ the size of the alphabet. A visualization of the algorithm can be found on \cite{web}.
\end{abstract}

\maketitle

\section{Introduction}
The word $u=abaaba$ satisfies $f(u)=u$ where $f$ maps $b$ to $aba$ and cancels $a$. Such words, which are fixed points of a nontrivial morphism, are called \emph{morphically imprimitive}. On the other hand, the word $u'=abba$ can be easily verified to be \emph{morphically primitive}, which means that the only morphism satisfying $f(u')=u'$ defined on $\{a,b\}^*$ is the identity.

Fixed points of word morphisms and morphically (im)primitive words are studied in \cite{head,HeadLando,shallit, daniel}. In \cite{DM}, the first polynomial algorithm is presented (called {\MF}) that decides whether a given word $w$ is morphically primitive. Moreover, given the input word $w$, it finds a corresponding morphism satisfying $f(w)=w$ with minimal number of letters mapped to a nonempty word (that is, not canceled).
 
The complexity of {\MF} is estimated  as $\bigo (m+\log n)\cdot n$ in \cite{DM}. Here we make more detailed analysis of the algorithm and improve the estimate to $\bigo(|E|\cdot n)$, where $E$ is the set of those letters $x$ for which $f(x)$ is nonempty.

\section{Definitions}
Let $\alp(w)$ denote the set of letters occurring in $w$ and $|w|$ the length of $w$. For a set $S\subset \alp(w)$,  denote by $|w|_S$ the number of all occurrences of letters from $S$ in $w$; we shorten $|w|_{\{a\}}$ as $|w|_a$. 

   Each morphism $f$, satisfying $f(w)=w$, induces a factorization of $w$, called  a \emph{morphic factorization}. 
 The morphic factorization consists of a set $E$ and a sequence $(w_1,w_2,\dots,w_k)$ such that
\begin{itemize}
	\item $w=w_1w_2\cdots w_k$,
	\item $|w_i|_E=1$ for each $i=1,2,\dots,k$, and
	\item if $|w_i|_e=|w_j|_e=1$ for some $e\in E$, then $w_i=w_j$.
\end{itemize}

 It is shown in \cite{head} that we can suppose, without loss of generality, that $f$ is idempotent, that is, $f(a)=f(f(a))$ for each $a\in \alp(w)$. (It is enough to iterate  a general $f$ sufficient number of times in order to obtain its idempotent version.)  Throughout the paper, we shall therefore assume that $f$ is idempotent. The relation between $f$ and the corresponding morphic factorization is then as follows:
for each $i=1,\dots,k$, we have $w_i=f(e)$, where $e$ is the unique letter from $E$ occurring in $w_i$, and $f(a)=\emp$ if $a\notin E$ (where $\emp$ denotes the empty word). Letters in $E$ are called \emph{expanding}.  We say that $E$ is a \emph{minimal set of expanding letters} if no proper subset of $E$ is the set of expanding letters for a morphism $f'$ satisfying $f'(w)=w$. In \cite{DM}, it is shown that all minimal sets of expanding letters have the same cardinality.
 
 Denote the $i$-the letter of $w$ by $w[i]$ and write $w[i\dots j]$, with $i\leq j$, to denote the factor $w[i]w[i+1]\cdots w[j]$ of $w$. We will also work with the set $\C_w$ of \emph{cuts}, that is, of borders between two consecutive letters (plus the beginning and the end of $w$). A word $w$ has $|w|+1$ cuts and we represent them by integers $0,1,\dots,|w|$. The cut $k$ is the border following the prefix of length $k$. Note that cuts $i,j$ delimit the factor $w[i+1\dots j]$. 

Given a word  $w$ and a morphism $f$ such that $f(w)=w$, we say that a cut $k$ is a \emph{left} cut if it lies in the image of an expanding letter on its left side. More formally, the cut $k$ is a left cut if $f(w[1\dots k])\leq k$.
Similarly, we say that $k$ is a \emph{right} cut if $f(w[1\dots k])\geq k$. Note that inequalities are not strict, therefore a cut $k$ can be both left and right, which happens if and only if $f(w[1\dots k])= k$. Note that cuts that are both left and right define the morphic factorization of $w$ induced by $f$.
We say that $w[i,j]$ is a \emph{stretch factor} if $i$ is a left cut and $j$ is a right cut. 

An important and natural notion is the \emph{neighborhood} of a letter $a$ in $w$, denoted by $\okol a$. The neighborhood of $a$ is the longest extension of $a$ which is possible for all occurrences of $a$ in $w$. It is easy to see that the word $\okol a$ contains exactly one occurrence of $a$, hence it can be written as $\okol a=\lev a a \prav a$.

We will need the following easy observation:
\begin{align}\label{okoli}
	\text{if $b\in \alp(\okol a)$, then $|w|_b\geq |w|_a$.}
\end{align}

Letters with minimal frequency in $w$ that occur in a given factor $u$ play a special role in the algorithm. Therefore, we define
$$\alpha(i,j)=\min\{k\tz i<k\leq j,\ |w|_{w[k]}\leq |w|_{w[k']}\ \text{for all}\ i<k'\leq j\}.$$
In other words, $\alpha(i,j)$ is the leftmost position of a least frequent letter in $w[i,j]$. Note that ``least frequent'' is measured with respect to whole $w$, not just with respect to $w[i,j]$.

\section{Description of the Algorithm}\label{popis}
The algorithm {\MF} is based on the following characterization of minimal expanding sets (for proofs and more details see \cite{DM}):

Let  $E$ be a minimal set of letters, and $L,R$ minimal sets of cuts satisfying the following 
\emph{stability conditions}:
\begin{enumerate}\renewcommand{\theenumi}{\Alph{enumi}}
	\item $\{0,|w|\}\subseteq L$, $\{0,|w|\}\subseteq R$. \label{sc1}
	\item \label{sc2} Let $w[k]=w[k']=a$ with $a\in E$, 
	Then 
					\begin{enumerate}
						\item $k-1 \in L$ and $k\in R$;\label{sc2a}
						\item $k+|\prav a|  \in L$ and $k-|\lev a|-1\in R$;\label{sc2b}
						\item for each $-|\lev a|-1\leq m\leq \prav a$ we have that \label{sc2c} 
											\begin{itemize}
												\item $k+m\in L$ if and only if $k'+m\in L$, and
												\item $k+m\in R$ if and only if $k'+m\in R$.  
											\end{itemize}
					\end{enumerate}
	\item If $i\in L$, $j\in R$ with $i<j$, then $w[\alpha(i,j)]\in E$. \label{sc3}		
\end{enumerate}
   Then $E$ is a minimal set of expanding letters. For $a\in E$, the image $f(a)$ is defined as 
   \[f(a)=\text{{\sc Image}}(a,E,L,R):= w[k-i,k+j],\]
 where $w[k]=a$; $i\geq 0$ is the smallest integer such that $k-i-1\in R$; and $j\geq 0$ is the largest integer such that 
\begin{itemize}
	\item $k+j\in R$, and
	\item $k+j'\notin L$ holds for  each $j'\leq j$.
\end{itemize}

Stability conditions guarantee that $f(a)$ is well defined, in particular, it is independent of the choice of $k$, and that the resulting morphism satisfies $f(w)=w$. Moreover, all cuts in $R$ are right cuts of the factorization, and cuts in $L$ are left cuts.
 In view of the fact that sets $E$, $L$ and $R$ represent expanding letters, left cuts and right cuts respectively, stability conditions can be rephrased informally as follows: 
\begin{enumerate}\renewcommand{\theenumi}{\Alph{enumi}}
 \item the extremal cuts are both left and right;
\item			\begin{enumerate}
				\item an expanding letter is delimited by a left and a right cut;
				\item neighborhood of an expanding letter is delimited by a right and a left cut (the left border is a right cut and vice versa);
				\item neighborhoods of expanding letters are synchronized with respect to left and right cuts;
			\end{enumerate}
	\item \label{inf3} the leftmost least frequent letter in each stretch factor is expanding.		
\end{enumerate}

The core procedure of the algorithm {\MF} consists in construction of sets $E$, $L$ and $R$ satisfying stability conditions. 
Given a subset $\tE$ of $\alp(w)$, we define subsets $\tL(\tE)$ and $\tR(\tE)$ of $\C_w$ as the smallest sets satisfying stability conditions \eqref{sc1} and \eqref{sc2}. Similarly, for two subsets $\tL$ and $\tR$ of $\C_w$, we define $\tE(\tL,\tR)$ as the smallest subset of $\alp(w)$ satisfying the stability condition \eqref{sc3}. We are looking for a set $\tE$ satisfying $\tE=\tE(\tL(\tE),\tR(\tE))$. If $\tE\neq \tE(\tL,\tR)$, then there exist cuts $i,j$ violating the condition  \eqref{sc3}, that is, the letter $w[\alpha(i,j)]$ is not an element of $E$. Denote such a letter by $\tN(\tE,\tL,\tR)$. The algorithm is now described by the following simple pseudocode.
\medskip
\begin{algorithm}{MorphicFactorization}{w}
		\tE\=\emptyset;\ \tL\=\{0,|w|\}; \tR\=\{0,|w|\}; \\
		\begin{WHILE}{\tE\neq \tE(\tL,\tR)} 
 \tE\=\tE \cup \{\tN(\tE,\tL,\tR)\};  \\
			\tL\=\tL(\tE);\ \tR\=\tR(\tE);
		\end{WHILE}\\
		\begin{FOR} {\text{each} a\in \alp(w)}
		\begin{IF}{a\in \tE} 
		{f(a)\=\CALL{Image}(a,\tE,\tL,\tR);}
		\ELSE
		{f(a)\=\emp;}
		\end{IF}
		\end{FOR}\\
		\RETURN{f};
\end{algorithm}
Several examples illustrating the work of the algorithm can be found in \cite{DM}. It can be also tested and visualized on \cite{web}. Here we add one more example. It can be also understood as a replacement of Example 7 in \cite{DM}, which is mistaken.

\begin{example}
Consider $w=caabcaadeaabeaad$, where $\okol a=a$, $\okol b=aab$, $\okol c=caa$,  $\okol d=aad$ and $\okol e=eaa$. Let us follow the run of the algorithm. At the beginning we set $\tE=\emptyset$ and $\tL=\tR=\{0,16\}$.
Rounds of the {\bf while} loop yield the following:
\begin{enumerate}
	\item[Round 1.]
\begin{itemize}
	\item	\eqref{sc3}	implies $\tN(\tE,\tL,\tR)=w[\alpha(0,16)]=w[1]=c$;
	\[
\begin{tikzpicture}[scale=0.7,inner sep=1pt]
\def\slovo{c/1,a/2,a/3,b/4,c/5,a/6,a/7,d/8,e/9,a/10,a/11,b/12,e/13,a/14,a/15,d/16}
\def\L{0,16}
\def\R{0,16}
\def\E{1,5}
\foreach \x in \E
\node[fill=black!20, inner sep=3.5pt, shape=circle] at (\x-0.5,0.5) {}; 
\draw[help lines] (0,0) grid (16,1);
\foreach \l/\p in \slovo
\node at (\p-0.5,0.5) {$\l$}; 
\foreach \x in {0,1,...,16}
{\draw[gray] node[fill=white] at (\x,0.5) {{\tiny \x}};
}
\foreach \x in \L
{\draw node[fill=white] at (\x,-0.3) {{\small \tt L}};
}
\foreach \x in \R
{\draw node[fill=white] at (\x,1.3) {{\small \tt R}};
}
\draw[gray] (0,0) .. controls (1,-1) and (2,-1) ..
node[fill=white] {\textcolor{black}{$\okol c$}}
(3,0);
\draw[gray] (4,0) .. controls (5,-1) and (6,-1) ..
node[fill=white] {\textcolor{black}{$\okol c$}}
(7,0);
\end{tikzpicture}
\]
	\item since $c\in \tE$, \eqref{sc2a} implies $0,4\in\tL$ , $1,5\in \tR$, and \eqref{sc2b} implies $3,7\in\tL$ , $0,4\in \tR$;
the condition \eqref{sc2c} is satisfied.
\end{itemize}
	
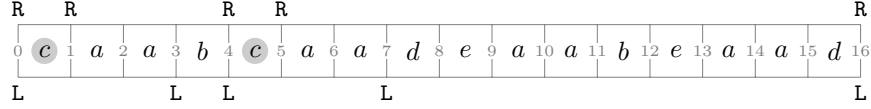
\[
\begin{tikzpicture}[scale=0.7,inner sep=1pt]
\def\slovo{c/1,a/2,a/3,b/4,c/5,a/6,a/7,d/8,e/9,a/10,a/11,b/12,e/13,a/14,a/15,d/16}
\def\L{0,3,4,7,16}
\def\R{0,1,4,5,16}
\def\E{1,5}
\foreach \x in \E
\node[fill=black!20, inner sep=3.5pt, shape=circle] at (\x-0.5,0.5) {}; 
\draw[help lines] (0,0) grid (16,1);
\foreach \l/\p in \slovo
\node at (\p-0.5,0.5) {$\l$}; 
\foreach \x in {0,1,...,16}
{\draw[gray] node[fill=white] at (\x,0.5) {{\tiny \x}};
}
\foreach \x in \L
{\draw node[fill=white] at (\x,-0.3) {{\small \tt L}};
}
\foreach \x in \R
{\draw node[fill=white] at (\x,1.3) {{\small \tt R}};
}
\end{tikzpicture}
\]
  \item[Round 2.]	
\begin{itemize}
	\item	\eqref{sc3}	implies $\tN(\tE,\tL,\tR)=w[\alpha(3,4)]=w[4]=b$;
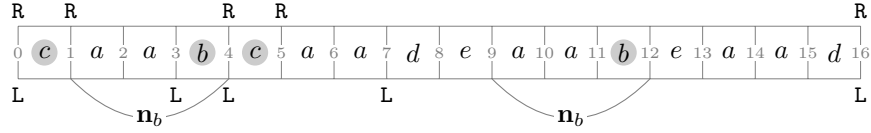
	\[
\begin{tikzpicture}[scale=0.7,inner sep=1pt]
\def\slovo{c/1,a/2,a/3,b/4,c/5,a/6,a/7,d/8,e/9,a/10,a/11,b/12,e/13,a/14,a/15,d/16}
\def\L{0,3,4,7,16}
\def\R{0,1,4,5,16}
\def\E{1,4,5,12}%
\foreach \x in \E
\node[fill=black!20, inner sep=3.5pt, shape=circle] at (\x-0.5,0.5) {}; 
\draw[help lines] (0,0) grid (16,1);
\foreach \l/\p in \slovo
\node at (\p-0.5,0.5) {$\l$}; 
\foreach \x in {0,1,...,16}
{\draw[gray] node[fill=white] at (\x,0.5) {{\tiny \x}};
}
\foreach \x in \L
{\draw node[fill=white] at (\x,-0.3) {{\small \tt L}};
}
\foreach \x in \R
{\draw node[fill=white] at (\x,1.3) {{\small \tt R}};
}
\draw[gray] (1,0) .. controls (2,-1) and (3,-1) ..
node[fill=white] {\textcolor{black}{$\okol b$}}
(4,0);
\draw[gray] (9,0) .. controls (10,-1) and (11,-1) ..
node[fill=white] {\textcolor{black}{$\okol b$}}
(12,0);
\end{tikzpicture}
\]
	\item since $b\in \tE$, \eqref{sc2a} implies $3,11\in\tL$ , $4,12\in \tR$, and \eqref{sc2b} implies $4,12\in\tL$, $1,9\in \tR$;
the condition \eqref{sc2c} is satisfied.
	
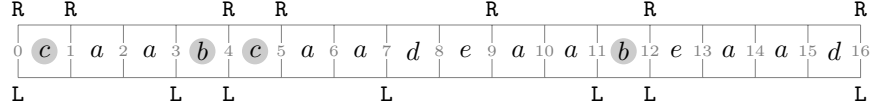
\[
\begin{tikzpicture}[scale=0.7,inner sep=1pt]
\def\slovo{c/1,a/2,a/3,b/4,c/5,a/6,a/7,d/8,e/9,a/10,a/11,b/12,e/13,a/14,a/15,d/16}
\def\L{0,3,4,7,11,12,16}
\def\R{0,1,4,5,9,12,16}
\def\E{1,4,5,12}%
\foreach \x in \E
\node[fill=black!20, inner sep=3.5pt, shape=circle] at (\x-0.5,0.5) {}; 
\draw[help lines] (0,0) grid (16,1);
\foreach \l/\p in \slovo
\node at (\p-0.5,0.5) {$\l$}; 
\foreach \x in {0,1,...,16}
{\draw[gray] node[fill=white] at (\x,0.5) {{\tiny \x}};
}
\foreach \x in \L
{\draw node[fill=white] at (\x,-0.3) {{\small \tt L}};
}
\foreach \x in \R
{\draw node[fill=white] at (\x,1.3) {{\small \tt R}};
}
\end{tikzpicture}
\]
 \end{itemize}
  \item[Round 3.]	
\begin{itemize}
	\item	\eqref{sc3}	implies $\tN(\tE,\tL,\tR)=w[\alpha(7,9)]=w[8]=d$;
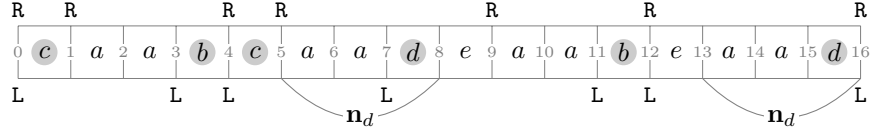
	\[
\begin{tikzpicture}[scale=0.7,inner sep=1pt]
\def\slovo{c/1,a/2,a/3,b/4,c/5,a/6,a/7,d/8,e/9,a/10,a/11,b/12,e/13,a/14,a/15,d/16}
\def\L{0,3,4,7,11,12,16}
\def\R{0,1,4,5,9,12,16}
\def\E{1,4,5,12,8,16}%
\foreach \x in \E
\node[fill=black!20, inner sep=3.5pt, shape=circle] at (\x-0.5,0.5) {}; 
\draw[help lines] (0,0) grid (16,1);
\foreach \l/\p in \slovo
\node at (\p-0.5,0.5) {$\l$}; 
\foreach \x in {0,1,...,16}
{\draw[gray] node[fill=white] at (\x,0.5) {{\tiny \x}};
}
\foreach \x in \L
{\draw node[fill=white] at (\x,-0.3) {{\small \tt L}};
}
\foreach \x in \R
{\draw node[fill=white] at (\x,1.3) {{\small \tt R}};
}
\draw[gray] (5,0) .. controls (6,-1) and (7,-1) ..
node[fill=white] {\textcolor{black}{$\okol d$}}
(8,0);
\draw[gray] (13,0) .. controls (14,-1) and (15,-1) ..
node[fill=white] {\textcolor{black}{$\okol d$}}
(16,0);
\end{tikzpicture}
\]
	\item since $d\in \tE$, \eqref{sc2a} implies $7,15\in\tL$ , $8,16\in \tR$, and \eqref{sc2b} implies $8,16\in\tL$, $5,12\in \tR$;
the condition \eqref{sc2c} is satisfied.
	
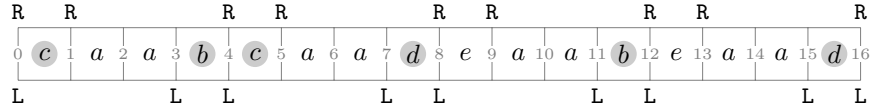
\[
\begin{tikzpicture}[scale=0.7,inner sep=1pt]
\def\slovo{c/1,a/2,a/3,b/4,c/5,a/6,a/7,d/8,e/9,a/10,a/11,b/12,e/13,a/14,a/15,d/16}
\def\L{0,3,4,7,8,11,12,15,16}
\def\R{0,1,4,5,8,9,12,13,16}
\def\E{1,4,5,12,8,16}%
\foreach \x in \E
\node[fill=black!20, inner sep=3.5pt, shape=circle] at (\x-0.5,0.5) {}; 
\draw[help lines] (0,0) grid (16,1);
\foreach \l/\p in \slovo
\node at (\p-0.5,0.5) {$\l$}; 
\foreach \x in {0,1,...,16}
{\draw[gray] node[fill=white] at (\x,0.5) {{\tiny \x}};
}
\foreach \x in \L
{\draw node[fill=white] at (\x,-0.3) {{\small \tt L}};
}
\foreach \x in \R
{\draw node[fill=white] at (\x,1.3) {{\small \tt R}};
}
\end{tikzpicture}
\]
 \end{itemize}
  \item[Round 4.]	
\begin{itemize}
	\item	\eqref{sc3}	implies $\tN(\tE,\tL,\tR)=w[\alpha(8,9)]=w[9]=e$;
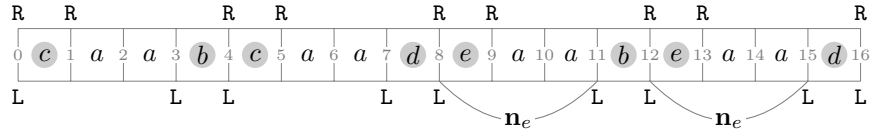
	\[
\begin{tikzpicture}[scale=0.7,inner sep=1pt]
\def\slovo{c/1,a/2,a/3,b/4,c/5,a/6,a/7,d/8,e/9,a/10,a/11,b/12,e/13,a/14,a/15,d/16}
\def\L{0,3,4,7,8,11,12,15,16}
\def\R{0,1,4,5,8,9,12,13,16}
\def\E{1,4,5,12,8,16,9,13}
\foreach \x in \E
\node[fill=black!20, inner sep=3.5pt, shape=circle] at (\x-0.5,0.5) {}; 
\draw[help lines] (0,0) grid (16,1);
\foreach \l/\p in \slovo
\node at (\p-0.5,0.5) {$\l$}; 
\foreach \x in {0,1,...,16}
{\draw[gray] node[fill=white] at (\x,0.5) {{\tiny \x}};
}
\foreach \x in \L
{\draw node[fill=white] at (\x,-0.3) {{\small \tt L}};
}
\foreach \x in \R
{\draw node[fill=white] at (\x,1.3) {{\small \tt R}};
}
\draw[gray] (8,0) .. controls (9,-1) and (10,-1) ..
node[fill=white] {\textcolor{black}{$\okol e$}}
(11,0);
\draw[gray] (12,0) .. controls (13,-1) and (14,-1) ..
node[fill=white] {\textcolor{black}{$\okol e$}}
(15,0);
\end{tikzpicture}
\]
	\item all conditions \eqref{sc2} are satisfied.
 \end{itemize}
\end{enumerate}
The remaining part of the algorithm {\sc MorphicFactorization} defines 
\[f: a\mapsto \emp, \quad b\mapsto aab,\quad c\mapsto c,\quad d\mapsto aad,\quad e\mapsto e.\] 
Note that also
\[f: a\mapsto \emp, \quad b\mapsto ab,\quad c\mapsto ca,\quad d\mapsto ad,\quad e\mapsto ea,\] 
and
\[f: a\mapsto \emp, \quad b\mapsto b,\quad c\mapsto caa,\quad d\mapsto d,\quad e\mapsto eaa\] 
are possible morphisms with the same set of expanding letters.
\end{example} 


\section{Complexity analysis}
In this section, we show that the complexity of the algorithm is in $\bigo(m\cdot n)$, where $n$ is the length of the analyzed word, and $m$ is the number of its letters. More precisely, we show that the complexity is  in \[\bigo( |E|\cdot n),\] where $E$ is a minimal set of expanding letters.

The core of the algorithm is the {\bf while} loop. The condition $\tE=\tE(\tL,\tR)$ is checked $|E|+1$ times and the loop is performed $|E|$ times since in each round one letter is added to $\tE$. Therefore, we have to prove that each round of the loop can be performed in $\bigo(n)$. 

It is convenient to calculate, during the initialization phase, the value of $|w|_a$ for each $a\in \alp(w)$, and also an array ${\Pos}[a,i]$, which yields the position of the $i$-th occurrence of $a$ in $w$.  The initialization phase is linear: it is enough to read the input once.

\subsection{Evaluation of the loop condition}
Evaluation of the loop condition consists in checking whether the stability condition \eqref{sc3} is satisfied. If it is not, then the evaluation also outputs the letter $\tN(\tE,\tL,\tR)$. This is done as follows. 

Look through cuts ${\tt l}$ in $\tL$  in increasing order and for each $\tl$ find the smallest cut ${\tr}\in \tR$ strictly larger than $\tl$, and ${\tk}=\alpha(\tl,\tr)$. If $w[{\tt k}]\notin \tE$, then we have found $\tN(\tE,\tL,\tR)$ and start the next round of the {\bf while} loop. If ${\tt l}=n$ and no violation of \eqref{sc3} was detected, return $\tE=\tE(\tL,\tR)$.

Note that ${\tt r}$ and ${\tt k}$ can never decrease, therefore the procedure is in $\bigo(n)$. However, not all factors $w[i,j]$ with $i\in\tL$ and $j\in\tR$ are checked; hence it has to be to shown that the stability condition \eqref{sc3} is verified correctly. 

Suppose, for a contradiction, that our procedure outputs $\tE=\tE(\tL,\tR)$, although $i\in \tL$ and $j\in \tR$ violate the the stability condition \eqref{sc3}. Assume that $j-i$ is as small as possible. Let $j'<j$ be the smallest cut in $\tR$ strictly larger than $i$ and let $k'=\alpha(i,j')$.
Since the stretch factor $w[i,j']$ has been checked by the procedure, we deduce $w[k']\in \tE$ (and $k'=j'$). 
On the other hand, by assumption, we have $w[k]\notin \tE$, where $k=\alpha(i,j)$. Hence $k'<k$ and $|w|_{k}<|w|_{k'}$. The stability condition \eqref{sc2b} implies $i'=k'+|\prav{{w[k']}}|\in \tL$, and we deduce $i'<k$, since the letter $w[k]$ is not in $\okol{w[k']}$ by \eqref{okoli} . 
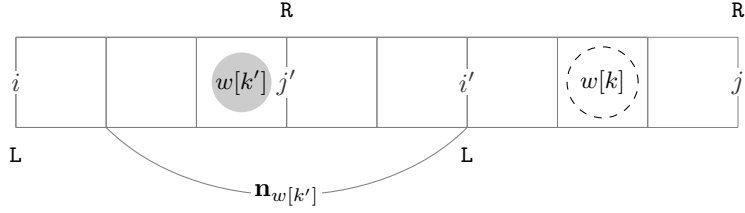
\[
\begin{tikzpicture}[scale=1.2, inner sep=1pt]
\draw[help lines] (0,0) grid (7,1);
\def\L{0,5}
\def\R{3,8}
\def\E{1,4,5,12,8,16}%
\draw[help lines] (0,0) grid (8,1);
\foreach \x/\y in {0/i,3/j',5/i',8/j}
{\draw[black!80] node[fill=white] at (\x,0.5) {$\y$};
}
\foreach \x in \L
{\draw node[fill=white] at (\x,-0.3) {{\small \tt L}};
}
\foreach \x in \R
{\draw node[fill=white] at (\x,1.3) {{\small \tt R}};
}
\node[fill=black!20, inner sep=8pt, shape=circle] at (3-0.5,0.5) {}; 
\node at (3-0.5,0.5) {\mbox{\small $w[k']$}};
\node[inner sep=9.5pt, shape=circle, draw, dashed] at (7-0.5,0.5) {}; 
\node at (7-0.5,0.5) {\mbox{\small $w[k]$}}; 
\draw[gray] (1,0) .. controls (2,-1) and (4,-1) ..
node[fill=white] {\textcolor{black}{$\okol {w[k']}$}}
(5,0);
\end{tikzpicture}
\]
Clearly, $k=\alpha(i,j)=\alpha(i',j)$, whence the factor $w[i',j]$ violates \eqref{sc3} too, a contradiction with minimality of $j-i$.

\subsection{Construction of {\tL} and \tR}
 The construction of sets $\tL$ and $\tR$ in each round consists in checking the stability condition {\eqref{sc2}} (the stability condition $\eqref{sc1}$ is fullfilled by the first line of the algorithm).  

 The condition \eqref{sc2a} says that, for a new letter $a\in \tE$,  we have to add positions immediately before occurrences of $a$ to the set $\tL$, and positions immediately after its occurrences to the set $\tR$. This can be done in $\mathcal O(|w|_a)$. 
\medskip

Similarly, the condition \eqref{sc2b} adds starting positions of $\okol a$ to $\tR$, and ending positions to $\tL$, where $a$ is a letter newly added to $\tE$. This requires to calculate $\okol a$, which is done as follows. In order to calculate $|\prav a|$, check, for growing $k\geq 1$, whether all letters
\[w[{\Pos}[a,i]+k], \quad i=1,2,\dots,|w|_a\]
agree, until a mismatch is encountered for $k=|\prav a|+1$. Similarly, with decreasing $k\leq -1$, it is possible to calculate $|\lev a|$. The
notion of a neighborhood implies that neighborhoods of different occurrences of the same letter cannot overlap too much; each position lies in at most two distinct neighborhoods of the same letter: once in its left part and
once in its right part. The number of positions visited during the calculation is therefore at most $2n$. We conclude that the cost of calculating $\okol a$ and of satisfying \eqref{sc2b} is in $\bigo(n)$. 
\medskip

 The stability condition \eqref{sc2c} is the most complex one. It can be concisely described as keeping all neighborhoods $\okol a$ of the same letter $a$ from $\tE$  synchronized.
	The underlying structure is an undirected graph with vertices $\C_w$ satisfying the following condition:
\[\text{cuts $\Pos[a,i]+k$ and $\Pos[a,i']+k$}\] 
are connected for each \[\text{$a\in \tE$, $1\leq i,i'\leq |w|_a$ and $-|\lev a|-1\leq k\leq |\prav a|$.} \]	
	 The condition \eqref{sc2c} then requires that connected cuts either all are, or all are not elements of $\tL$ (of $\tR$ resp.). In other words, being in $\tL$ (in $\tR$ resp.) is a property of a connected component rather than of an individual cut. 
	We shall represent this information as a forest of rooted trees of height one. Each cut is linked to its parent, which is the root representing the connected component. The root also keeps the information whether the component is in sets $\tL$, $\tR$. Checking whether the cut is in $\tL$ (in $\tR$ resp.) therefore requires constant time. 

 When a new letter $a$ is added to $\tE$, new edges synchronizing neighborhoods of $a$ have to be added too, and the graph becomes more complex. To satisfy the condition \eqref{sc2c} as it is formulated in the previous paragraph,  it is enough to add 
edges
\[ (\Pos[a,1]+k,\Pos[a,i]+k) \]
for $i=2,\dots,|w|_a$ and $-|\lev a|-1\leq k\leq |\prav a|$.
The number of new edges can be bounded by an argument similar to the one used above when calculating neighborhoods: each cut is the second vertex of a new edge at most two times. This implies that the number of new edges is less than $2n$.    
 After new edges have been added, the algorithm searches the whole graph and compresses the connected components back to the forest of height one. Since the graph has at most $n$ old vertices and at most $2n$ new ones, this can be done in $\mathcal O(n)$.

The final definition of $f$ is clearly in $\bigo (n)$, which completes the proof. 

\section{Conclusion}
We have shown that morphic primitivity can be tested in linear time for fixed alphabet. This may be surprising compared with the fact that a similar problem, checking the existence of a morphism between two \emph{distinct} words, is NP-complete (cf. \cite{Ehrenfeucht}).

If the alphabet is not fixed, the algorithm is at worst quadratic, consider for example the family of morphically primitive words \[w_n=a_1a_2\cdots a_{n-1}a_na_na_{n-1}\cdots a_2a_1,\]
for which the main loop of the algorithm runs $n/2$ rounds. On the other hand, our analysis implies that it can be checked in linear time that all letters in $w_n$ have trivial neighborhoods, whence the morphic primitivity follows. 
Precise complexity in the uniform case therefore remains unclear.

\end{document}